\documentclass[twocolumn,showpacs,preprintnumbers,amsmath,amssymb]{revtex4}

\usepackage{graphicx}
\usepackage{dcolumn}
\usepackage{bm}
\usepackage{color}

\begin{document}

\title{An explanation for the pseudogap of high-temperature superconductors based on quantum optics}

\author{Deshui Yu}
\author{Jingbiao Chen}\thanks{jbchen@pku.edu.cn}

\affiliation{Institute of Quantum Electronics, School of Electronics
Engineering and Computer Science, Peking University, Beijing 100871,
P. R. China}

\begin{abstract}
We first explain the pseudogap of high-temperature superconductivity
based on an approach of quantum optics. After introducing a damping
factor for the lifetime $\tau$ of quasiparticles, the
superconducting dome is naturally produced, and the pseudogap is the
consequence of pairing with damped coherence. We derive a new
expression of Ginzburg-Landau free energy density, in which a
six-order term due to decoherence damping effect is included.
Without invoking any microscopic pairing mechanism, this approach
provides a simple universal equation of second-order phase
transition, which can be reduced to two well-known empirical scaling
equations: the superconducting dome Presland-Tallon equation, and
the normal-state pseudogap crossover temperature $T^{*}$ line.
\end{abstract}


\maketitle
Besides the unveiled microscopic mechanism, the
remarkable mysteries of high-temperature superconductivity (HTSC)
are the superconducting dome, the HTSC transition temperature
$T_{c}$ as a function of doping, and pseudogap in phase
diagram~\cite{Millis,Emery1,Norman1}. It is believed that, a correct
explanation for the superconducting dome and pseudogap is strongly
related to the origin of HTSC. However, there is no convincing
explanation with detailed analytical expressions for the
superconducting dome and pseudogap region. Its origin and physical
interpretation are fundamental questions in HTSC
~\cite{Millis,Emery1,Norman1}. There are two most popular physical
models on HTSC: Resonating-valence-bond (RVB)
picture~\cite{Anderson,Lee} and quantum critical
scenario~\cite{Tallon,Norman2}. The former origins from the idea of
spin singlet scenario~\cite{Nagaosa} and that the pseudogap is a
precursor to the superconducting state, which is pairing without
long-range phase coherence~\cite{Emery2}. The latter origins form
the idea that another state of matter, pseudogap phase, competes
with superconducting gap for the same Fermi surface.

In this paper,we first apply quantum optics approach to explain
HTSC, and show that the experimental results of superconducting dome
together with pseudogap can be described by a universal equation of
second-order phase transition. The superconducting dome and
normal-state pseudogap crossover temperature in HTSC are attributed
to the finite lifetime of cooper pairs due to decoherence damping,
which results in localized cooper pairs with limited phase
coherence. We also present a modified general Ginzburg-Landau free
energy density (GLFED), which includes a sixth-order term due to
this decoherence damping effect. This phenomenological two-state
model does not invoke any microscopic pairing mechanism, but can
provide a simple universal equation, which can be reduced to two
well-known empirical scaling equations: the superconducting dome
Presland-Tallon equation~\cite{Presland}, and the normal-state
pseudogap crossover temperature $T^{*}$ line~\cite{Huefner}.

Before applying quantum optics approach, we should list some
commonly accepted facts we ensure for HTSC: (1) Superconductivity is
due to the formation of Cooper pairs with long-range
order~\cite{Tsuei,Harlingen}, which are spin singlet and have
$d_{x^{2}-y^{2}}$ orbital symmetry. (2) Photoemission experiments
reveal sharp spectral peaks in the excitation
spectrm~\cite{Damascelli}, indicating the presence of quasiparticle
states. (3) The essential structure is the CuO$_{2}$ planes, and the
formation of Cooper pairs takes place independently within different
multiplayers~\cite{Dagotto}. (4) The electron-phonon interaction is
not the principal mechanism of the formation of Cooper
pairs~\cite{Anderson2}. (5) The Cooper pairs are formed from the
time-reversed states (called as non-pairing
breaking)~\cite{Leggett}.

First we show the BCS Hamiltonian can be expressed in quantum optics
approach. The Hamiltonian of the whole system can be expressed as
$H=\sum_{i=0}^{4}H_{i}$, where
\begin{equation}\label{}
    H_{0}=\sum_{\mathbf{p},\sigma,\mathrm{i}}\left(\varepsilon_{\mathbf{p}}-\mu\right)c^{+}_{\mathbf{p},\sigma,\mathrm{i}}c_{\mathbf{p},\sigma,\mathrm{i}}+\sum_{\mathbf{q}}\hbar\omega_{\mathbf{q}}a^{+}_{\mathbf{q}}a_{\mathbf{q}},
\end{equation}
is the Hamiltonian of the free system, the Hamiltonian of the
interaction between quasiparticles and intermediate bosons is
\begin{equation}\label{}
    H_{1}=\sum'_{\mathbf{p},\mathbf{p}',\sigma,\mathrm{i}}M_{\mathbf{p}',\mathbf{p}}\left(c^{+}_{\mathbf{p}',\sigma,\mathrm{i}}c_{\mathbf{p},\sigma,\mathrm{i}}a_{\mathbf{p}'-\mathbf{p}}+H.c\right),
\end{equation}
the Hamiltonian of the repulsive interaction between quasiparticle
and quasiparticle is
\begin{equation}\label{}
    H_{2}=\sum'_{\mathbf{p},\mathbf{p}',\mathrm{i}}V_{\mathbf{p}',\mathbf{p}}c^{+}_{\mathbf{p}',\uparrow,\mathrm{i}}c^{+}_{-\mathbf{p}',\downarrow,\mathrm{i}}c_{-\mathbf{p},\downarrow,\mathrm{i}}c_{\mathbf{p},\uparrow,\mathrm{i}},
\end{equation}
and the coupling Hamiltonian between neighboring CuO$_{2}$ layers is
\begin{equation}\label{}
    H_{3}=\sum'_{\mathbf{p},\sigma,\mathrm{i}}U\left(c^{+}_{\mathbf{p},\sigma,\mathrm{i}}c_{\mathbf{p},\sigma,\mathrm{i}+1}+c^{+}_{\mathbf{p},\sigma,\mathrm{i}+1}c_{\mathbf{p},\sigma,\mathrm{i}}\right),
\end{equation}
and the random Hamiltonian
$H_{4}=\sum'_{\mathbf{p},\sigma,\mathrm{i}}h^{R}_{\mathbf{p},\sigma,\mathrm{i}}$.
$H_{4}$ includes all the random interaction, such as the in-elastic
interaction, the random potential caused by doping, and so on. This
Hamiltonian could introduce a damping rate of quasiparticles. Here,
$\sum'_{\mathbf{p},\mathbf{p}',\mathrm{i}}$ denotes the sum of
quasiparticles that participate in Cooper pairing.
$c^{+}_{\mathbf{p},\sigma,\mathrm{i}}$ creates a Fermi-type
quasiparticle with momentum state $\mathbf{p}$=($p_{x}$, $p_{y}$),
spin $\sigma$ in the $i$th CuO$_{2}$ layer. In the low-temperature
superconductors, the Cooper pair formed by two electrons is caused
by the electron-phonon interaction. It is not true in HTSC as the
facts we list above. However, there must be the intermediate bosons
being changed among the quasiparticles in order to induce the
attractive potential between two quasiparticles. Here we define
$a_{\mathbf{q}}$ as the annihilation operator of the intermediate
boson with momentum $\mathbf{q}$=($q_{x}$, $q_{y}$).
$M_{\mathbf{p}',\mathbf{p}}$ is the quasiparticle-boson interaction
strength, $V_{\mathbf{p}',\mathbf{p}}$ is the
quasiparticle-quasiparticle interaction strength and $U$ is the
coupling strength between two neighboring layers. \emph{$H_{1}$ is
the energy of the quasiparticle-quasiparticle repulsive interaction,
while $H_{2}$ is the energy of quasiparticle-boson interaction,
which could cause Cooper pairing. Here, we ignore the effect of
$H_{3}$. If $|H_{2}|>|H_{1}|$, the repulsive interaction will
suppress Cooper pairing, for which no superconductivity occurs.
However, if $|H_{2}|<|H_{1}|$, many Cooper pairs could be formed,
which will leads superconductivity.}

In conventional BCS theory, the Hamiltonian of the electron-phonon
interaction $H_{1}$ is included in the electron-electron interaction
$H_{2}$ resulting an effective attractive potential. However, in
HTSC, since there is no consensus on the dispersion of quasiparticle
and the $d$-type Fermi surface, it is difficult to follow the
conventional BCS approach to investigate HTSC. In this case, we
include the $H_{2}$ in $H_{1}$, and the system Hamiltonian is given
by $H=H_{0}+V+H_{3}+H_{4}$, where
$V=\sum'_{\mathbf{p},\mathbf{p}',\sigma,\mathrm{i}}h_{\mathbf{p},\mathbf{p}',\sigma,\mathrm{i}}$
is the effective quasiparticle-boson interaction, with
\begin{equation}\label{}
    h_{\mathbf{p},\mathbf{p}',\sigma,\mathrm{i}}=M_{\mathbf{p}',\mathbf{p}}\left(c^{+}_{\mathbf{p}',\sigma,\mathrm{i}}c_{\mathbf{p},\sigma,\mathrm{i}}A_{\mathbf{p}'-\mathbf{p},\bar{\sigma}}+H.c\right),
\end{equation}
and the effective boson operator
$A_{\mathbf{p}'-\mathbf{p},\bar{\sigma}}$ is
\begin{equation}\label{}
    A_{\mathbf{p}'-\mathbf{p},\bar{\sigma}}=a_{\mathbf{p}'-\mathbf{p}}+\frac{1}{4}\frac{V_{\mathbf{p}',\mathbf{p}}}{M_{\mathbf{p}',\mathbf{p}}}c^{+}_{-\mathbf{p}',\bar{\sigma},\mathrm{i}}c_{-\mathbf{p},\bar{\sigma},\mathrm{i}},
\end{equation}
where $\bar{\sigma}$ denotes the opposite $\sigma$. Since the most
important physics is included in $H_{0}+V+H_{4}$ and the use of
$H_{3}$ is only to ensure the long-range order (there is no
long-range order in the two-dimensional system), we only consider
the interaction of the two-dimensional system, firstly.

\begin{figure}
\includegraphics[width=8cm]{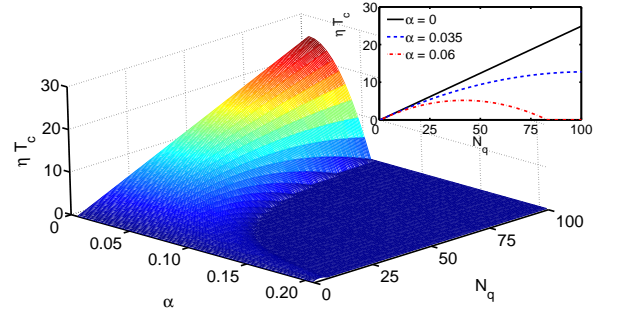}\\
\caption{The temperature $\eta T_{c}$ as a function of parameter
$\alpha$ and density of quasiparticles $N_{q}$ with zero
longitudinal damping $\beta=0$ and $\Delta T=0$. For zero $\alpha$,
temperature $T$ is proportional to the density of quasiparticles
$N_{q}$. However, in the non-zero $\alpha$ case, the dependence of
temperature on $N_{q}$ is arc shape, and dome shape for larger
$\alpha$, as shown in the inserted firgure.} \label{Figure2}
\end{figure}

The interaction Hamiltonian $V$ is composed by many two-state
systems, in which $h_{\mathbf{p},\mathbf{p}',\sigma,\mathrm{i}}$
denotes that the two states of quasiparticle,
$|e\rangle=|\mathbf{p}',\sigma,\mathrm{i}\rangle$ and
$|g\rangle=|\mathbf{p},\sigma,\mathrm{i}\rangle$ are coupled by a
boson with moment $\mathbf{q}=\mathbf{p}'-\mathbf{p}$. The number of
intermediate bosons with momentum $\mathbf{q}$ is $\left\langle
A^{+}_{\mathbf{p}'-\mathbf{p},\bar{\sigma}}A_{\mathbf{p}'-\mathbf{p},\bar{\sigma}}\right\rangle$.
Here we should emphasize that since $V_{\mathbf{p}',\mathbf{p}}>0$
is the repulsive potential and quasiparticle-boson interaction leads
an attractive potential, we have
\begin{equation}\label{}
\left\langle
A^{+}_{\mathbf{p}'-\mathbf{p},\bar{\sigma}}A_{\mathbf{p}'-\mathbf{p},\bar{\sigma}}\right\rangle<\left\langle
a^{+}_{\mathbf{p}'-\mathbf{p}}a_{\mathbf{p}'-\mathbf{p}}\right\rangle.
\end{equation}
which denotes that there is a bias for the Cooper pairing. That is
to say only the number of bosons larger than a certain value can
Cooper pairing occurs. It will be seen below.
$h_{\mathbf{p},\mathbf{p}',\sigma,\mathrm{i}}+h^{R}_{\mathbf{p},\sigma,\mathrm{i}}+h^{R}_{\mathbf{p}',\sigma,\mathrm{i}}$
can be simply expressed as
$h_{e,g}=h^{R}+M_{e,g}\left(c^{+}_{e}c_{g}A_{eg}+H.c\right)$. This
model has been well investigated in Quantum Optics~\cite{Scully}. In
Heisenberg picture, $\langle c^{+}_{e,g}(t)c_{e,g}(t)\rangle$ give
the number of quasiparticles in states $|e,g\rangle$ as a function
of time, and $\langle c^{+}_{g}(t)c_{e}(t)\rangle$ denotes the
coherence between two quasiparticle states. On the other hand, in
the conventional BCS theory, the lifetime quasiparticle is infinite
(or very long). However, in HTSC, the random Hamiltonian gives a
finite lifetime $\tau=\gamma^{-1}$ to quasiparticle since for
arbitrary operator $Q(t)$ we have
$\frac{d}{dt}Q(t)=\frac{1}{i\hbar}[Q(t),h^{R}]=-\gamma Q(t)$. The
damping rate $\gamma$ includes all the decoherent factors that could
destroy the coherence of Cooper pairing and could be expressed as a
polynome of the density of quasiparticles, which participate in
Cooper pairing, $N_{q}$ (defined below),
$\gamma=\Sigma_{j}a_{j}(N_{q})^{j}$. Here we only consider the first
two terms, thus $\gamma$ can be written as
$\gamma=\gamma_{T}+\gamma_{L}$, where the transverse damping rate
$\gamma_{T}=\tau^{-1}_{T}$, which may be caused by spin-exchange
collision~\cite{Wittke}, is proportional to $N_{q}$, and
$\gamma_{L}=\tau^{-1}_{L}$ denotes the longitudinal damping rate,
which is unrelated to $N_{q}$, but relate to the
superconductivity-transition temperature $T_{c}$.

Considering the statistic, we define the quantum average variables
as $\phi_{\alpha\beta}(t)\equiv\langle
c^{+}_{\alpha}(t)c_{\beta}(t)\rangle$. In Heisenberg picture, the
dynamic equation for operator $Q(t)$ is
$i\hbar\frac{d}{dt}Q(t)=[Q(t),h_{\mathbf{p},\mathbf{p}',\sigma,\mathrm{i}}+h^{R}_{\mathbf{p},\sigma,\mathrm{i}}+h^{R}_{\mathbf{p}',\sigma,\mathrm{i}}]$,
and
$i\hbar\frac{d}{dt}Q(t)=[Q(t),h_{\mathbf{p},\mathbf{p}',\sigma,\mathrm{i}}]-i\hbar\gamma
Q(t)$. Thus, the dynamics of simplified system could be expressed as
\begin{eqnarray}
  \frac{d}{dt}\phi_{ee} &=& -\gamma\phi_{ee}+i\Omega\phi_{ge}-i\Omega^{*}\phi_{eg}, \\
  \frac{d}{dt}\phi_{gg} &=& -\gamma\phi_{gg}-i\Omega\phi_{ge}+i\Omega^{*}\phi_{eg}, \\
  \frac{d}{dt}\phi_{eg} &=&
  -\gamma\phi_{eg}-i\Omega\left(\phi_{ee}-\phi_{gg}\right),
\end{eqnarray}
where $\Omega=\frac{M_{eg}}{\hbar}\sqrt{n_{bos}-n_{nor}}$ is the
general Rabi frequency. We have used the fact that $A_{eg}\approx
A^{+}_{eg}\approx\sqrt{n_{bos}-n_{nor}}$ since
$(n_{bos}-n_{nor})\gg1$. $n_{bos}$ is the number of bosons with
moment $\mathbf{q}=\mathbf{p}'-\mathbf{p}$, and $n_{nor}$ is the
equivalent boson number caused by the quasiparticle-quasiparticle
repulsive potential. Thus, one could see that the number of bosons
should be larger than the bias number $n_{nor}$. If there is no
damping factor, $\gamma=0$, the perfect Rabi oscillation of between
states $|e,g\rangle$ is presented, which is given by
$\phi_{eg}(t)=\frac{i}{2}\sin2|\Omega|t$. However, if $\gamma\neq0$,
$\phi_{eg}(t)$ should be averaged by the lifetime of quasiparticle
$\tau$ as $\bar{\phi}_{eg}=\frac{1}{\tau}\int
dt\phi_{eg}(t)e^{-t/\tau}$, which leads to an expression of the
second-order phase transition~\cite{Scully}
\begin{equation}\label{}
\bar{\phi}_{eg}=\frac{i}{2}\frac{|\Omega|^{2}}{|\Omega|^{2}+\tau^{-2}}.
\end{equation}
We define that the quasiparticle density of two states are $n_{e}$
and $n_{g}$, and their sum $n_{e}+n_{g}=n_{q}$ is the density of
quasiparticles with momentum $\mathbf{p}'$ and $\mathbf{p}$. Since
$|\bar{\phi}_{eg}|$ denotes the average value of transition
probability, which is also equal with $|\bar{\phi}_{ge}|$,
$n_{e}|\bar{\phi}_{eg}|$ gives the number of bosons radiated by the
upper state $|e\rangle$, and $n_{g}|\bar{\phi}_{ge}|$ gives the
number of bosons absorbed by the lower state $|g\rangle$. By summing
all momentum parts and spin, we have that
$N_{q}=\sum'_{\mathbf{p},\mathbf{p}',\sigma}n_{q}$ is the total
density of quasiparticles that participate in Cooper pairing,
$N_{bos}=\sum'_{\mathbf{p},\mathbf{p}',\sigma}n_{bos}$ denotes the
total density of intermediate bosons, and
$N_{nor}=\sum'_{\mathbf{p},\mathbf{p}',\sigma}n_{nor}$ is the
effective bosen density caused by the repulsive interaction.

Since the transverse damping rate $\tau^{-1}_{T}$ is proportional to
the number of quasiparticles~\cite{Wittke} while $\tau^{-1}_{L}$
does not, thus $\frac{1}{g\tau_{T}}=\alpha N_{q}$ and
$\beta=\frac{1}{g\tau_{L}}$, with $g=\frac{|M_{eg}|}{\hbar}$.
$\alpha$ is a proportional parameter, and $\beta$ will be zero for
zero traverse damping rate. In the equilibrium state, we have
$n_{g}=n_{e}=\frac{1}{2}n_{q}$, and
$n_{e}|\bar{\phi}_{eg}|=n_{bos}-n_{nor}$. Thus, the relation between
temperature $T$ and the density of quasiparticles $N_{q}$ could be
expressed by an equation of second-order phase transition,
\begin{equation}\label{EquilibriumEquation}
\eta
(T-T_{c})=\alpha^{2}N^{2}_{q}-\left(\frac{1}{4}-2\alpha\beta\right)N_{q}+\eta
\Delta T.
\end{equation}
Here we have used the fact that
$N_{nor}-N_{bos}-\beta^{2}=\eta(T-T_{c}-\Delta T)$, where $\eta$ a
proportional parameter to make equal dimension on the two sides of
equal sign. This relation is based on the well-known
Gingzburg-Landau free energy density (GLFED, see
below)~\cite{Scully}. $\eta\Delta T$ is a temperature bias. $T_{c}$
is the critical temperature, below which ($T<T_{c}$) the materials
are in the superconducting state. \emph{As we have said, $N_{bos}$
is the boson number contributing the attractive potential, while
$N_{nor}$ is the effective boson number contributing the repulsive
potential. Thus, if $N_{bos}<N_{nor}$ ($T>T_c$), the repulsive
potential will suppress the attractive potential. No
superconductivity occurs in this case. However, the attractive
potential will suppress the repulsive potential for
$N_{bos}>N_{nor}$ ($T<T_c$), which will leads superconductivity
transition.} In this case, we could get the condition for materials
being in the superconducting state from
Eq.(\ref{EquilibriumEquation}):
\begin{equation}\label{}
0\leq-\alpha^{2}N^{2}_{q}+\left(\frac{1}{4}-2\alpha\beta\right)N_{q}-\eta
\Delta T\leq\eta T_{c},
\end{equation}
and the critical condition of the superconductivity transition
\begin{equation}\label{CriticalEquilibriumEquation}
\eta
T_{c}=-\alpha^{2}N^{2}_{q}+\left(\frac{1}{4}-2\alpha\beta\right)N_{q}-\eta
\Delta T.
\end{equation}
Figure 1 shows the temperature $T_{c}$ as a function of $N_{q}$ for
different parameter $\alpha$ with zero longitudinal damping rate
$\beta=0$ and $\Delta T=0$. One could see that, for zero transverse
damping rate, $\alpha=0$, the density of quasiparticles $N_{q}$ is
proportional to the temperature $T_{c}$. However, for the non-zero
$\alpha$ case, the dependence between $T_{c}$ and $N_{q}$ is arc
shape, and finally, dome shape. It is very like the pseudogap in the
diagram of HTSC, in which $T^{\ast}$ is proportional to the doping,
and the dependence between $T_{c}$ and doping is a dome shape. The
detailed discussion will be shown in below. By the way, Fig.1
coincides with the results of $\mu$SR measurements in
Ref.~\cite{Uemura}.

Equation (\ref{EquilibriumEquation}) is the steady result for the
system. We could treat it as the consequence of the minimum of GLFED
$f_{s}$, if we define an order parameter $\psi_{i}$, of which
$|\psi_{i}|^{2}$ gives the density of quasiparticles $N_{q}$. In
this case GLFED could be expressed as
\begin{eqnarray}\label{GLFED}
\nonumber f_{s} &=& f_{n}+\left(A|\psi_{i}|^{2}+B|\psi_{i}|^{4}+C|\psi_{i}|^{6}\right) \\
&& +\frac{U}{N}\sum_{i}\left(\psi_{i}\psi^{*}_{i+1}+c.c\right),
\end{eqnarray}
where $A=\eta\left(-\Delta T-T_{c}+T\right)/2$,
$B=\left(\frac{1}{4}-2\alpha\beta\right)/4$ and $C=-\alpha^{2}/6$,
and $N$ is the total number of CuO$_{2}$ layers. Below we will see
that in the superconductive phase, $\Delta T$ is negative, and
$\frac{1}{4}\gg2\alpha\beta$. $f_{n}$ is the free energy density of
the normal state. As one could see, unlike the usual
GLFED~\cite{Chakravarty}, the sixth-order term of $|\psi|$, which is
completely caused by the damping factors of quasiparticles, is
included. We think the the coupling interaction between two
neighboring layers less contributes to the free energy. $\Delta T$
is a temperature bias. In the special case, $\alpha=0$, which
denotes the lifetime of quasiparticle is infinite, Eq.(\ref{GLFED})
will come back to the usual form of GLFED.

From above GLFED one could discuss the sample face energy and the
response of HTSC in external magnetic field. Here we should denote
that: Several phenomenal theories~\cite{Nagaosa,Sachdev} based on
the Ginzburg-Landau equations have been established to explain the
pseudogap and superconductivity dome of HTSC. However, these
theories have two disadvantages: I. One order parameter corresponds
to one phase. Thus what one should find is the order parameter
corresponding to the superconductivity phase, but not those for
fermion pairing and the Bose condensation in RVB picture or the
competing order in quantum critical scenario. II. Pseudogap and the
$d$-density wave (DDW) have not been confirmed to be a phase. There
is no reason to introduce an order parameter to either.

\begin{figure}
\includegraphics[width=9cm]{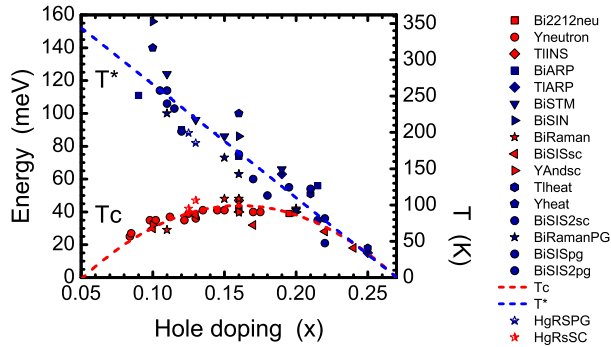}\\
\caption{Energy of pseudogap (blue points) and superconducting gap
(red points) for a number of HTSCs as measured as a function of hole
doping $x$ (obtained from Ref.~\cite{Huefner}). Lines are the
consequences derived from our theory with the parameters
$\eta^{-1}=152$ meV, $\alpha=0.27$, $\eta\Delta T=-0.13$, $y=0.24$,
$\beta=0.065$ and $\mu=18$. Blue line gives the temperature of
pseudogap $T^{\ast}$ ($\alpha=0$), while red line denotes the
superconductivity-transition temperature $T_{c}$.} \label{Figure4}
\end{figure}

In order to compare with the experiments, we should write
Eq.(\ref{CriticalEquilibriumEquation}) relating to the doping $x$.
Since $N_{q}$ must be a function of $x$, it could always be
expressed as a polynome of $x$, $N_{q}=\sum_{n}\lambda_{n}x^{n}$.
Here we only consider the first two terms, $N_{q}=\nu-\mu x$. Thus,
Eq.(\ref{CriticalEquilibriumEquation}) could be re-expressed as
\begin{equation}\label{pseudogap}
T_{c}=\eta^{-1}\left[-\alpha^{2}\mu^{2}(y-x)^{2}+\left(\frac{1}{4}-2\alpha\beta\right)\mu(y-x)-\eta\Delta
T\right],
\end{equation}
where $y=\nu/\mu$. As one could see, for the "localized"
measurements, which do not include the transverse damping factor,
$\alpha=0$, the temperature is linear proportional to the doping
$x$. We believe it corresponds to the temperature $T^{\ast}$ of
pseudogap~\cite{Norman1}. On the other hand, in fact, quasiparticles
have finite lifetime, which leads non-zero $\alpha$. The dependence
between temperature and doping is arc shape, and dome shape for
lager parameter $\alpha$. In this case, Eq. (\ref{pseudogap})
simultaneously involves the well-known two empirical
equations~\cite{Presland, Huefner}:
$E_{pg}=E^{\textrm{max}}_{pg}(0.27-x)/0.22$,
$E_{sc}=E^{\textrm{max}}_{sc}\left[1-82.6(0.16-x)^{2}\right]$.
Figure \ref{Figure4} shows the comparison between our theoretical
result and the experimental datum of Ref.~\cite{Huefner}. They
perfectly match each other. We think that the experimental
techniques~\cite{Timusk}, such as angle-resolved photonmission
(ARP)~\cite{Damascelli} and nuclear magnetic resonance
(NMR)~\cite{Warren}, only detect the "localized" character of pairs
and do not include the effective transverse damping factor. This
predicts $T_{c}$ will increase with small transverse damping factor
$\alpha$.

From above discussion, one could see that the function of doping is
only to dilute the density of quasiparticles (or Cooper pairs) and
consequently to increase the effective lifetime of quasiparticle
(long-range coherence). In the low-doped region, the number of
quasiparticles is so large that their lifetimes are very short,
which leads the short-range order. At the optimal doping, the number
of Cooper pairs and their coherence match very well. In the
over-doped region, the number of Cooper pairs decrease, which
reduces the superconductivity $T_{c}$.

In conculion, we have applied an approach of quantum optics to
explain the pseudogap in HTSC. By introducing the effective lifetime
of quasiparticle, the superconducting dome is naturally produced. We
also derive a new expression of GLFED, which includes the six-order
term of order parameter and could simultaneously give the two
well-known empirical formulas~\cite{Presland, Huefner}. The main
results of this letter are the modified general GLFED expression Eq.
(6) and the universal equation of second-order phase transition Eq.
(8). Despite the simplicity of this approach, these general and
universal expressions should provide new insights into the origin of
HTSC.

This work is supported by MOST of China under Grant No.
2005CB724500.


\begin{thebibliography}{99}

\bibitem{Millis} A. J. Millis, Science
\textbf{314}, 1888 (2006).

\bibitem{Norman1} M. R. Norman, D. Pines, and C. Kallin, Adv. Phys. \textbf{54}, 715 (2005).

\bibitem{Emery1}
V. J. Emery, S. A. Kivelson and O. Zachar, Phys. Rev. B \textbf{56},
6120 (1997).

\bibitem{Anderson} P. W. Anderson, Science
\textbf{235}, 1196 (1987).

\bibitem{Lee} P. A. Lee, N. Nagaosa and Xiao-Gang Wen, Rev. Mod. Phys. \textbf{78}, 17 (2006).

\bibitem{Tallon}
J. L. Tallon \emph{et al.}, Phys. Stat. Sol. \textbf{215}, 531
(1999).

\bibitem{Norman2} M. R. Norman, and C. Pepin, Rep. Prog.
Phys. \textbf{66}, 1547 (2003).

\bibitem{Nagaosa} N. Nagaosa, and P.
A. Lee, Phys. Rev. B \textbf{45}, 966 (1992).

\bibitem{Emery2} V. J.
Emery, and S. A. Kivelson, Nature \textbf{374}, 434
(1995).

\bibitem{Presland} M. R. Presland, J.L.Tallon, R.G.Buckley,
R.S.Liu and N.E.Flower, Physica C \textbf{176}, 95
(1991).

\bibitem{Huefner} S. Huefner, M. A. Hossain, A. Damascelli,
and G. Sawatzky, Rep. Prog. Phys. \textbf{71}, 062501 (2008).

\bibitem{Tsuei} C. C. Tsuei and J. R. Kirtley, Rev. Mod.
Phys. \textbf{72}, 969 (2000).

\bibitem{Harlingen} Van Harlingen, D.
J. Rev. Mod. Phys. \textbf{67}, 515 (1995).

\bibitem{Damascelli} A.
Damascelli, Z. Hussain and Zhi-Xun Shen, Rev. Mod. Phys.
\textbf{75}, 473 (2003).

\bibitem{Dagotto} E. Dagotto, Rev. Mod. Phys. \textbf{66},
763 (1994).

\bibitem{Anderson2} P. W. Anderson, Science \textbf{288},
480 (2000).

\bibitem{Leggett} A. J. Leggett, Quantum
Liquids (Oxford Univ. Press, Oxford, 2006).

\bibitem{Uemura} Y. J.
Uemura, \emph{et al.}, Phys. Rev. Lett. \textbf{62}, 2317
(1989).

\bibitem{Scully} M. O. Scully and M. S. Zubairy,
\emph{Quantum Optics} (Cambridge Univ. Press, 1997), Chapter 6 and
11.

\bibitem{Wittke} J. P. Wittke, R. H. Dicke,
Phys. Rev.  \textbf{103}, 620 (1956).

\bibitem{Chakravarty} S. Chakravarty, \emph{et al.}, Nature
\textbf{428}, 53 (2004).

\bibitem{Sachdev} S.
Sachdev, Phys. Rev. B \textbf{45}, 389 (1992).

\bibitem{Timusk} T. Timusk,
and B. Statt, Rep. Prog. Phys. \textbf{62}, 61 (1999).

\bibitem{Warren} W. W.
J. Warren, \emph{et al.}, Phys. Rev. Lett. \textbf{62}, 1193
(1989).

\bibitem{Alloul} H. Alloul, T. Ohno, and P. Mendels, Phys. Rev.
Lett. \textbf{63}, 1700. (1989).

\end{thebibliography}

\end{document}